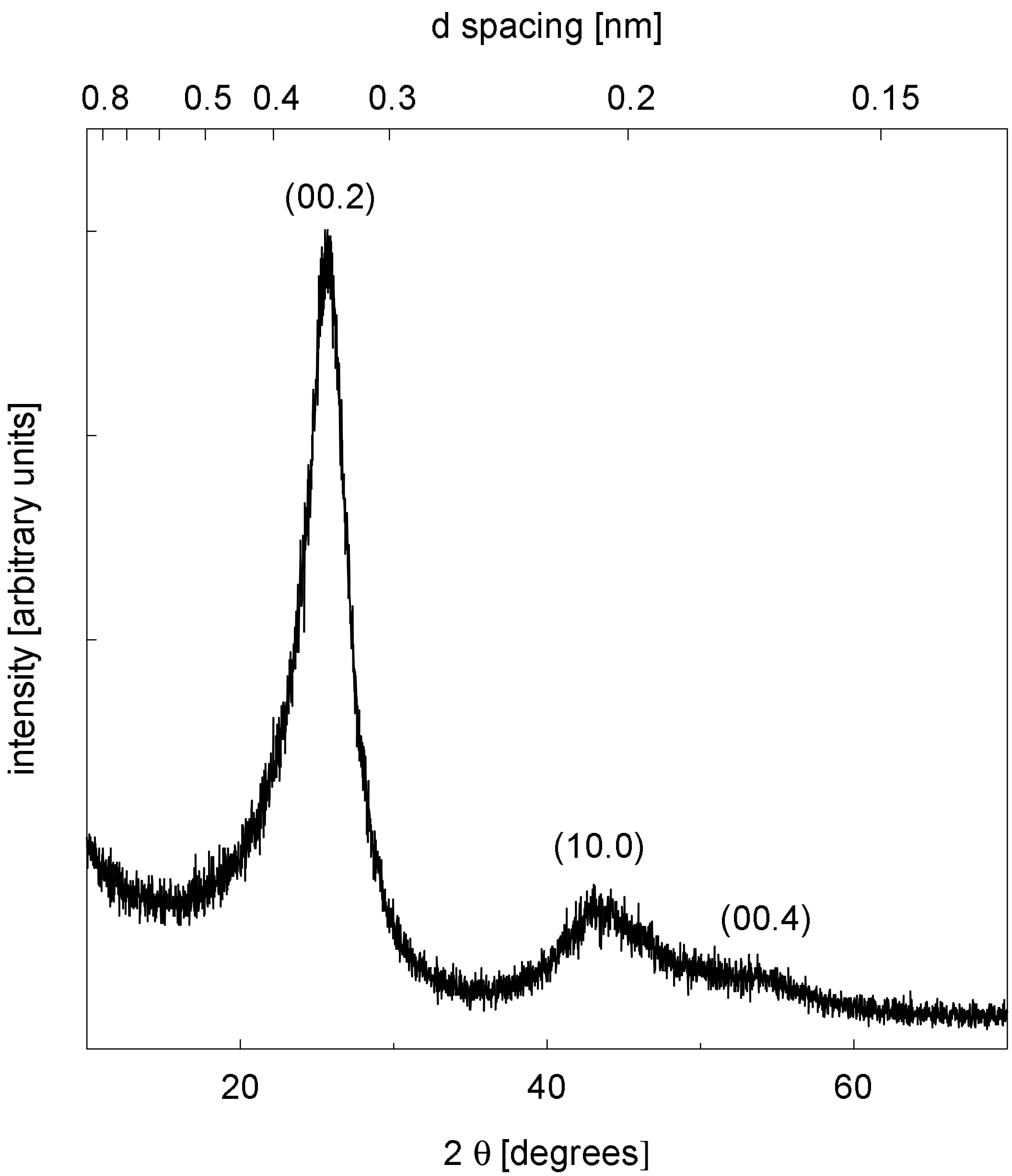
d spacing [nm]
0.8
0.5
0.4
0.3
0.2
0.15
(00.2)
(10.0)
(00.4)
intensity [arbitrary units]
20
40
60
2 θ [degrees]

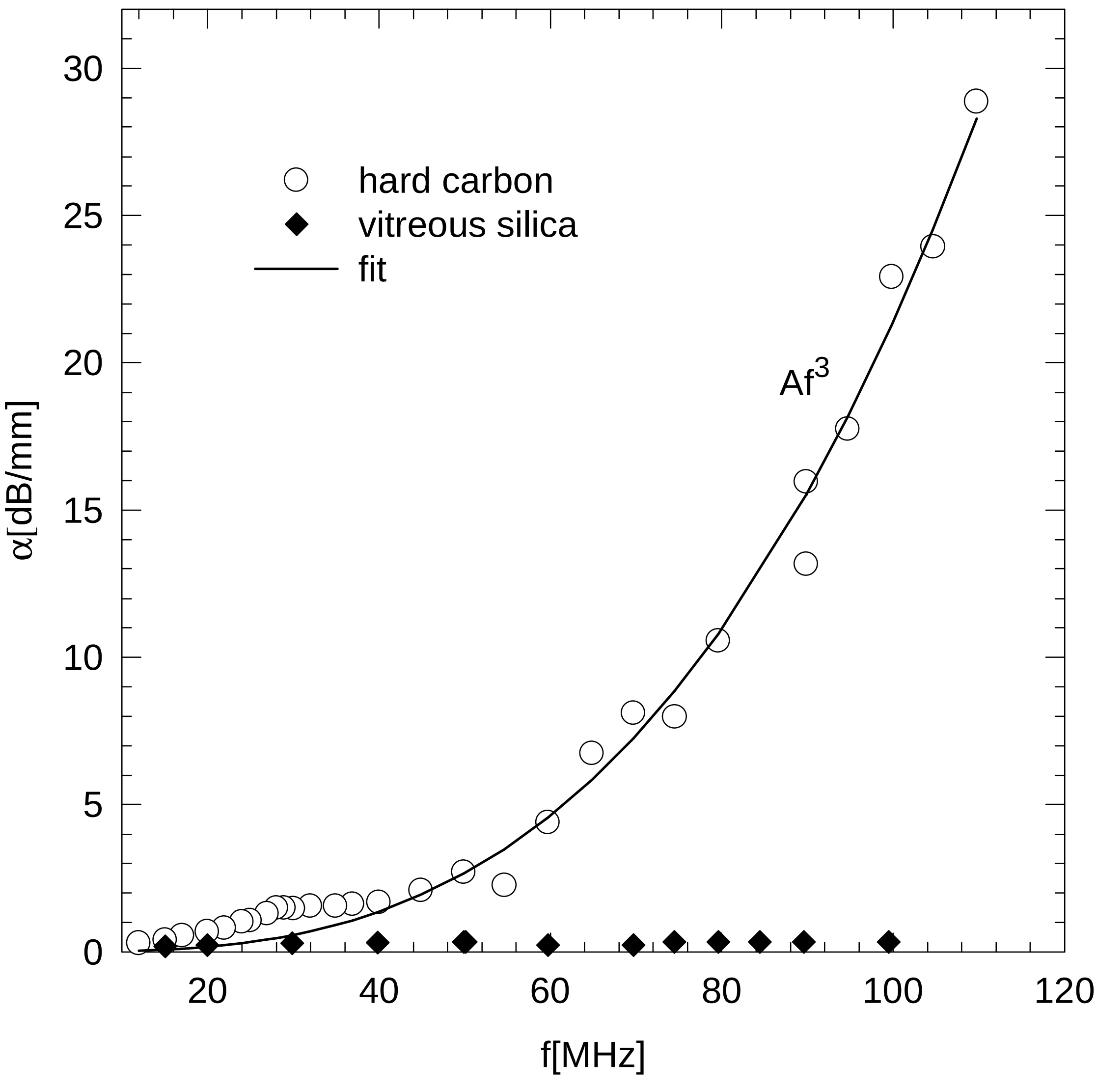

hard carbon
vitreous silica
fit
Af3
α[dB/mm]
f[MHz]
0
5
10
15
20
25
30
20
40
60
80
100
120

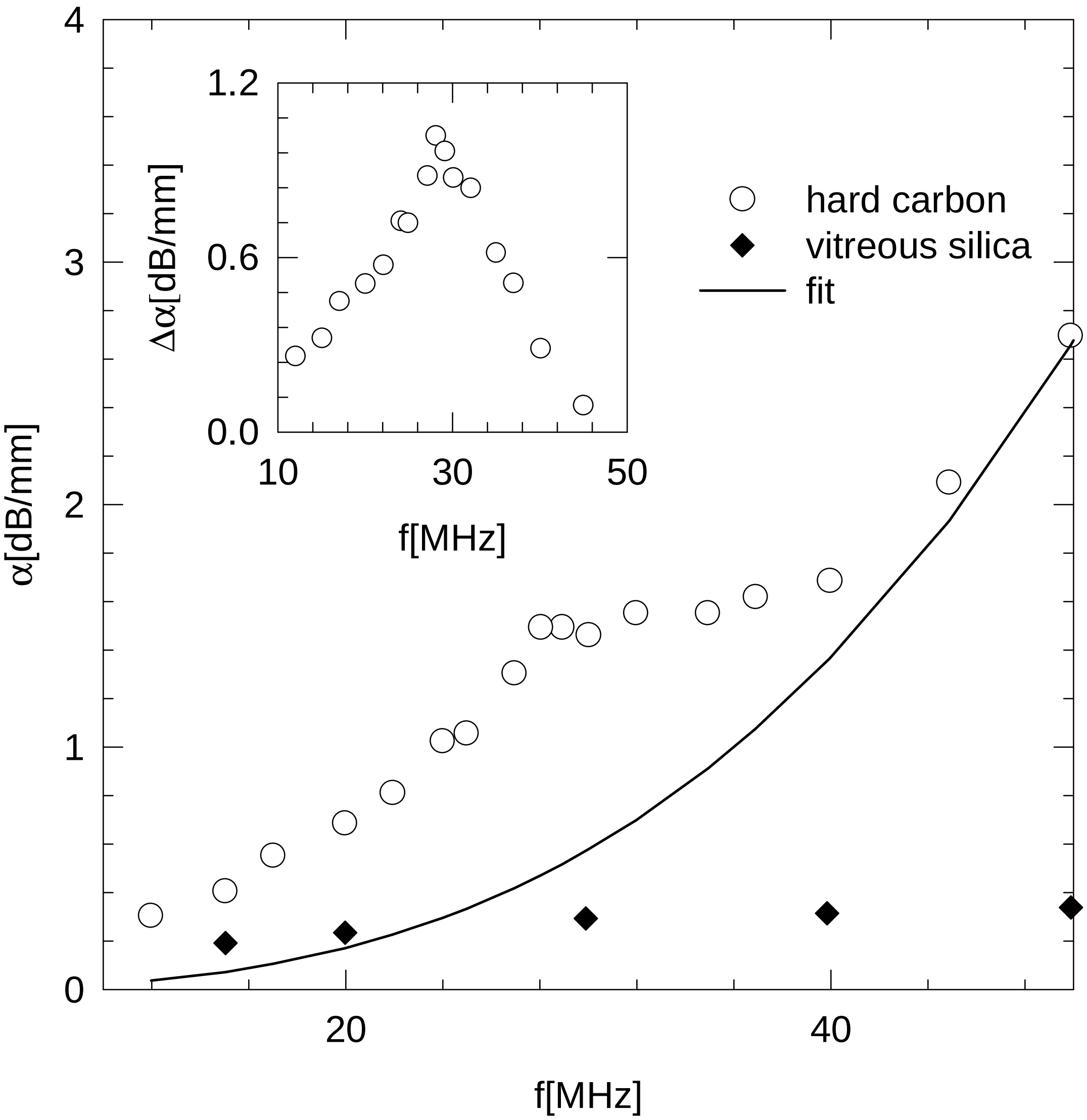
4
3
2
1
0
α[dB/mm]
20
40
f[MHz]
1.2
0.6
0.0
Δα[dB/mm]
10
30
50
f[MHz]
hard carbon
vitreous silica
fit

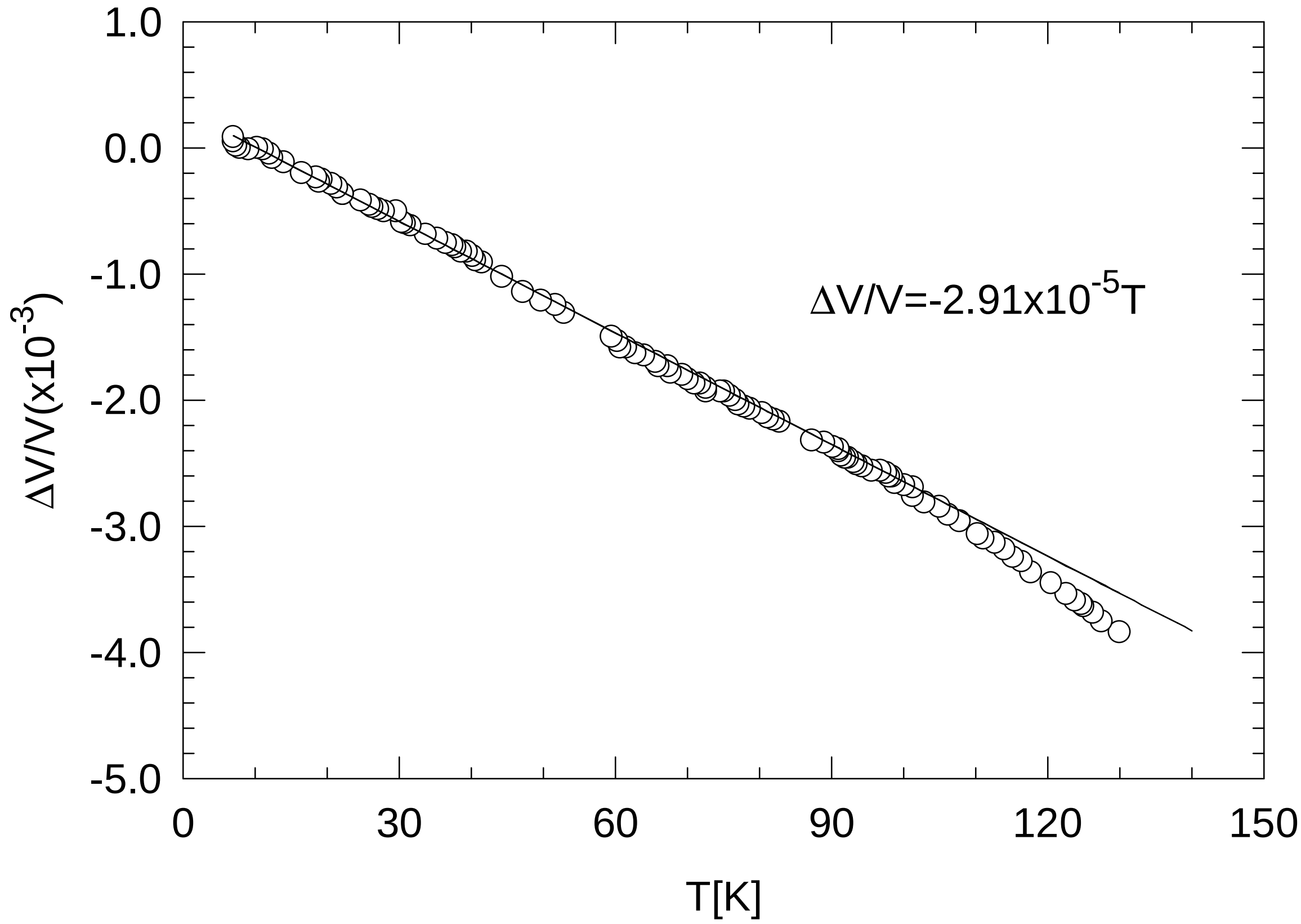

1.0
0.0
-1.0
-2.0
-3.0
-4.0
-5.0
0
30
60
90
120
150
T[K]
ΔV/V(x10$^{-3}$)
ΔV/V=-2.91x10$^{-5}$T

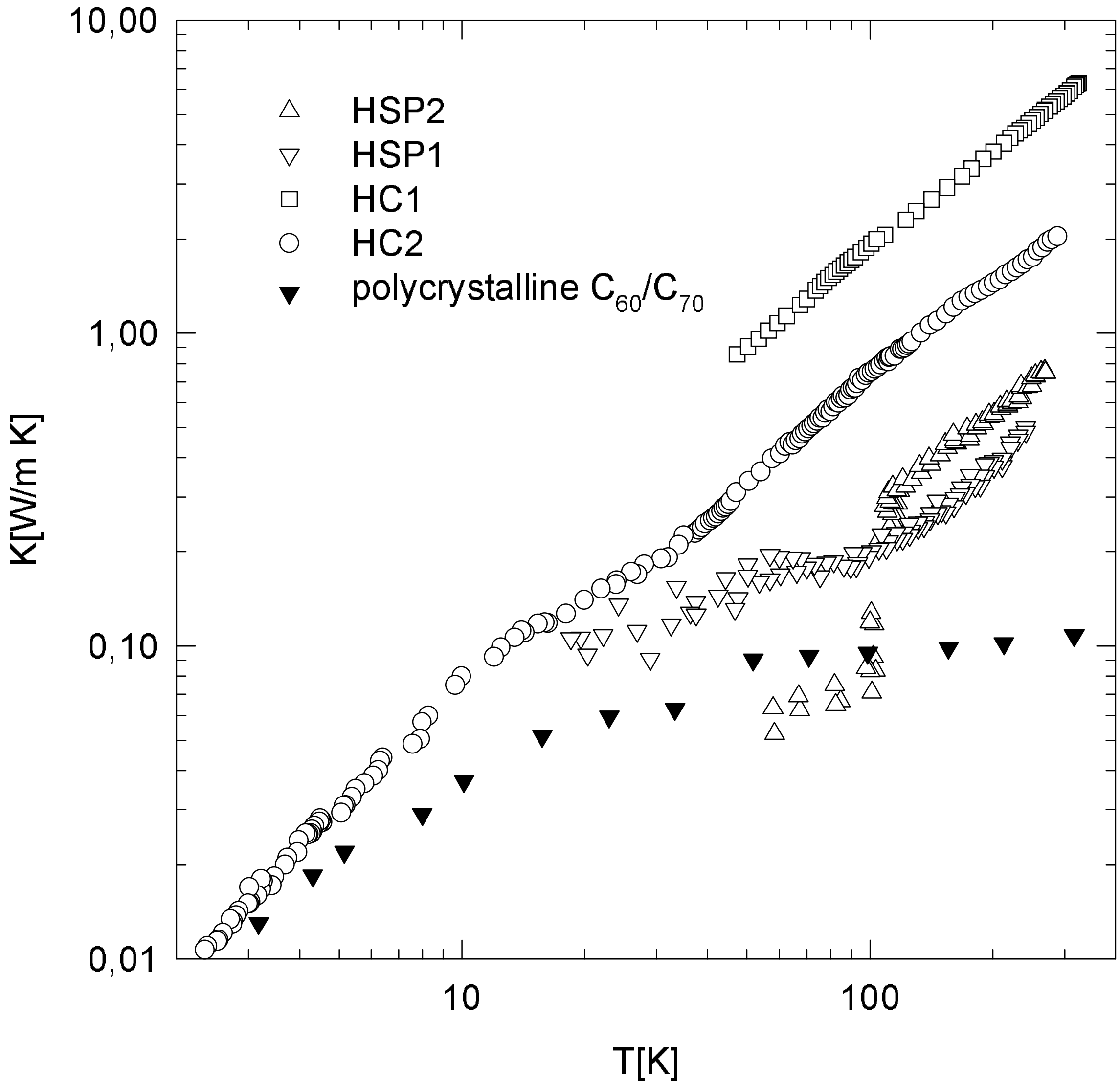

HSP2
HSP1
HC1
HC2
polycrystalline $C_{60}/C_{70}$
10,00
1,00
0,10
0,01
K[W/m K]
10
100
T[K]

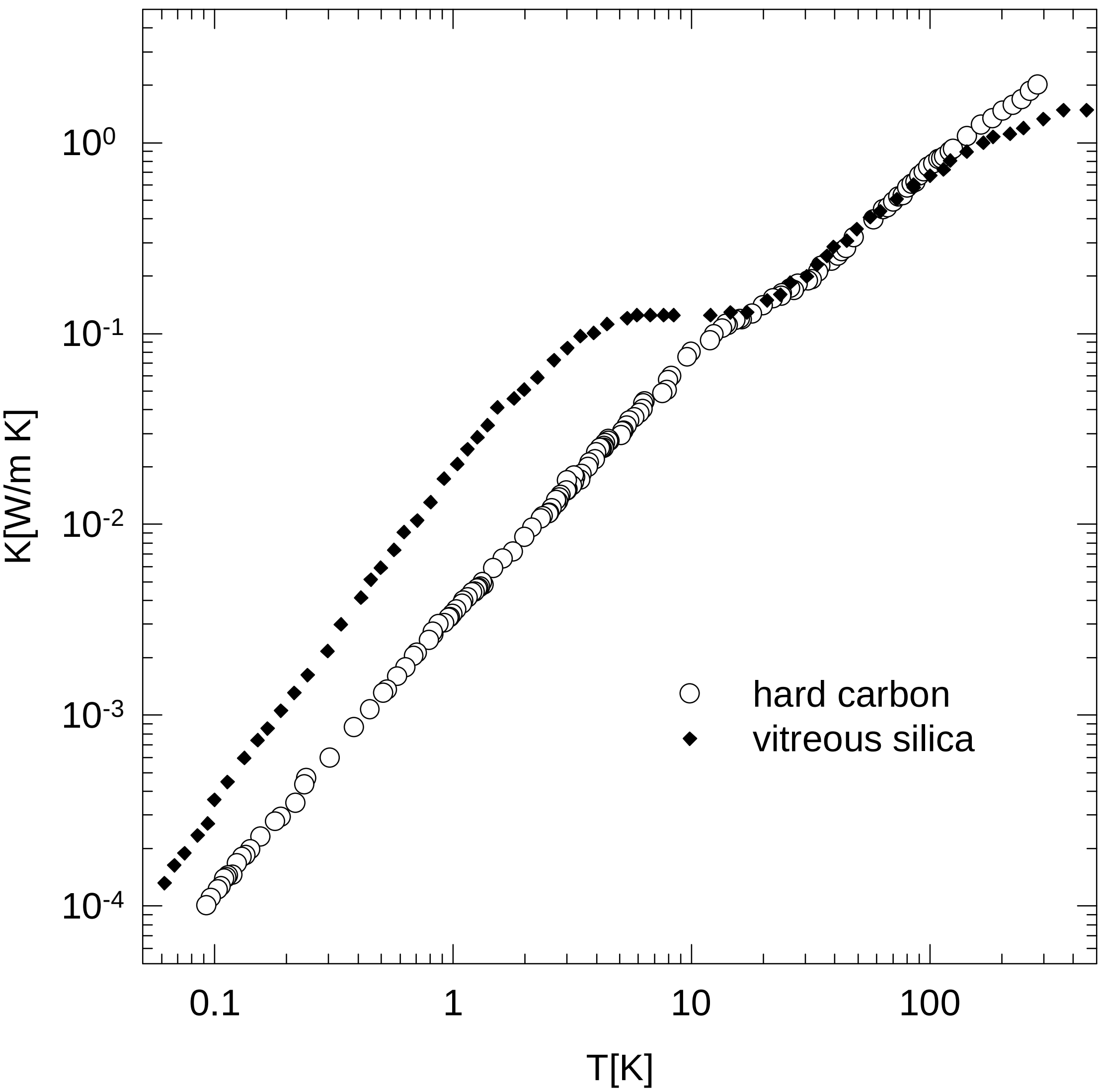

K[W/m K]
10^0
10^-1
10^-2
10^-3
10^-4
0.1
1
10
100
T[K]
hard carbon
vitreous silica

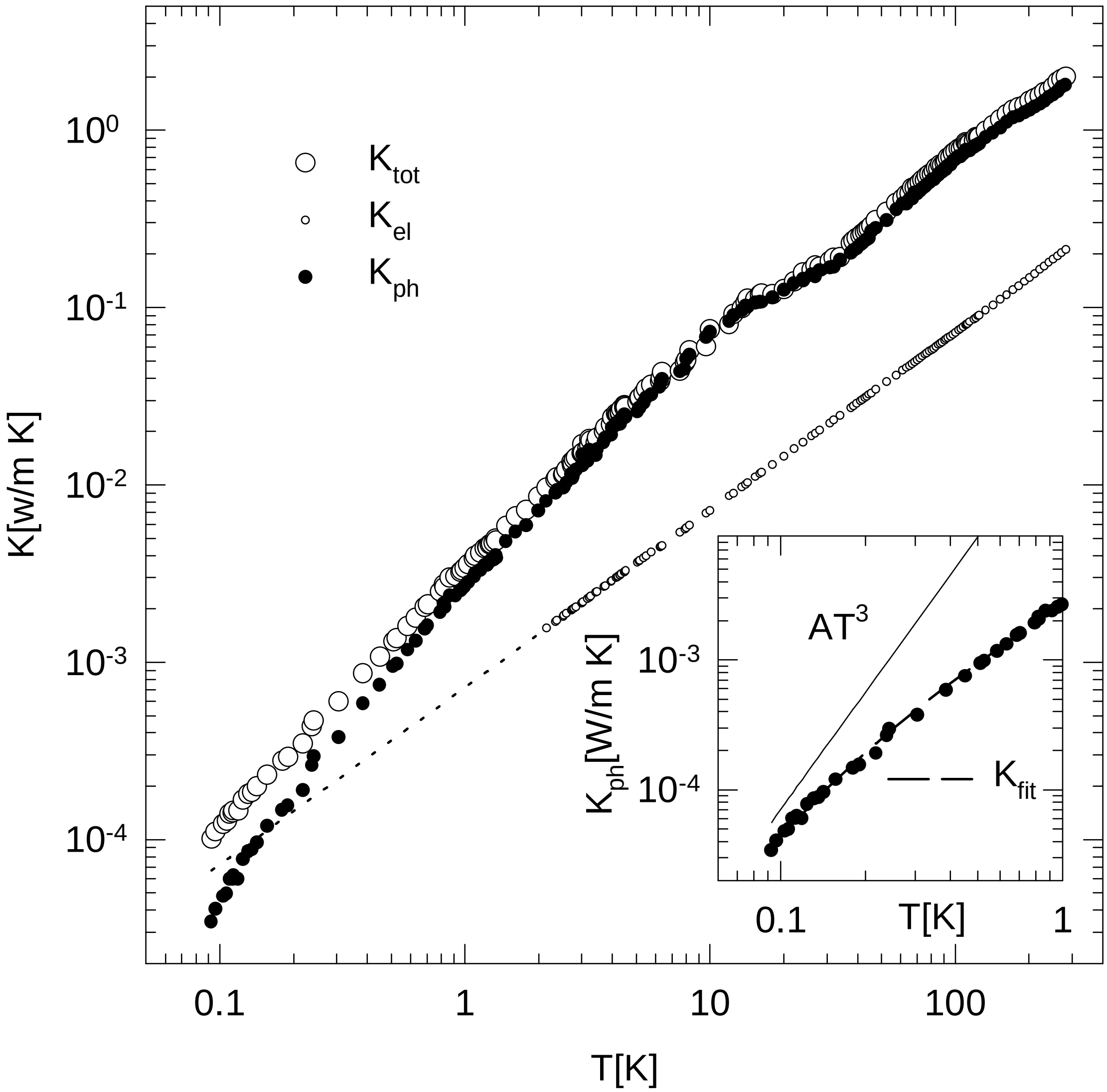

K_tot
K_el
K_ph
K[w/m K]
T[K]
AT^3
K_ph[W/m K]
K_fit

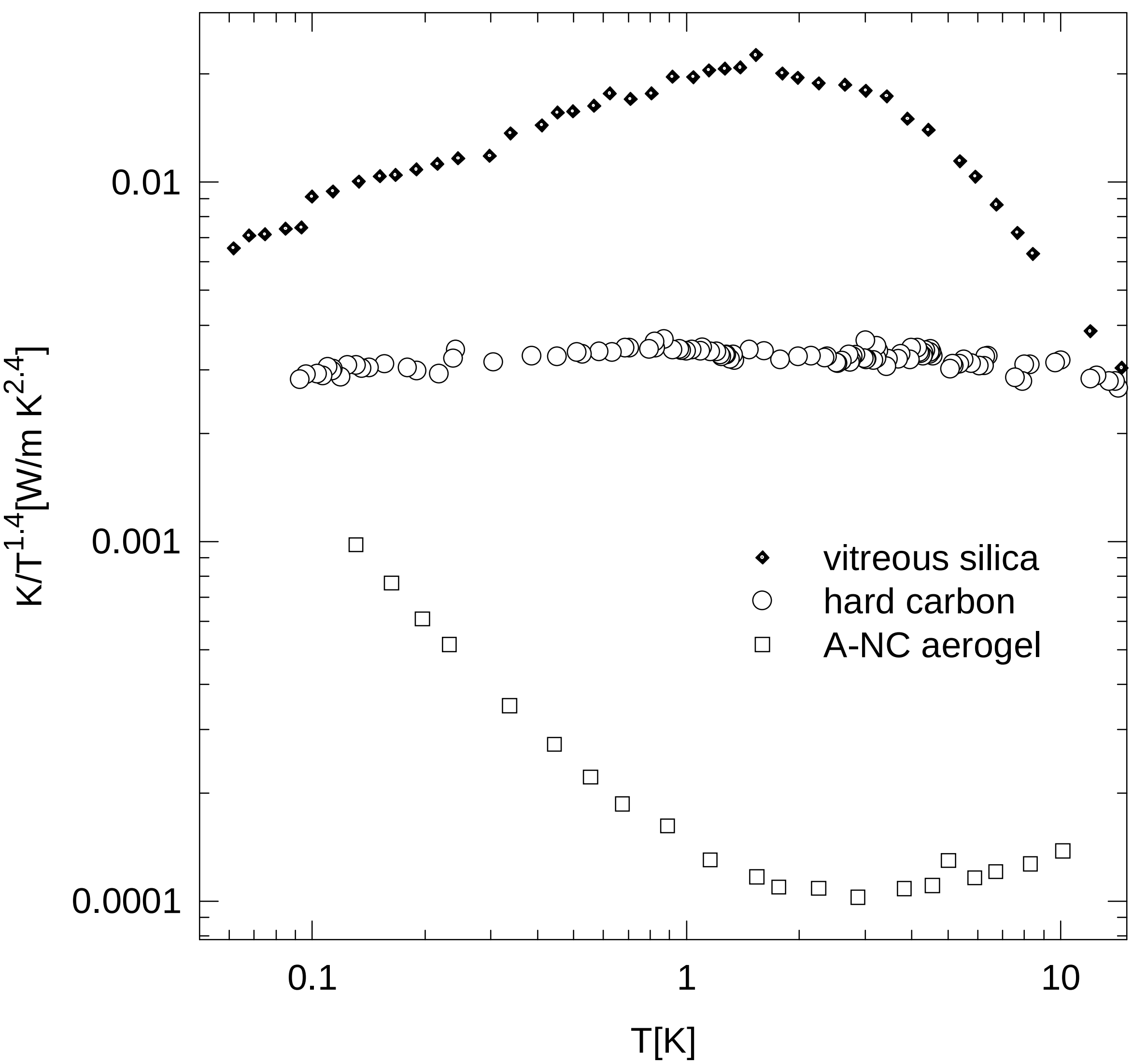

K/T^1.4[W/m K^2.4]
0.01
0.001
0.0001
0.1
1
10
T[K]
vitreous silica
hard carbon
A-NC aerogel